\documentstyle{elsart}
\include{epsf}

\begin{document}

\begin{frontmatter}
\title{Phase diagram and dependence of the critical \\
 temperature $T_c$ on the pressure  for
$Tl_{0.5}Pb_{0.5}Sr_2Ca_{1-x}Y_xCu_2O_7$ }

\author[UFF]{E. S. Caixeiro }, 
\author[UFF]{E. V. L. de Mello} 

\address[UFF]{Departamento de F\'{\i}sica,
Universidade Federal Fluminense, av. Litor\^ania s/n, Niter\'oi, R.J.,
24210-340, Brazil}
 
\thanks[caixa] {Partially supported by the Brazilian
agencies Capes, Faperj and CNPq.}

\begin{abstract}
Using a mean-field BCS-like approach on the bidimensional extended Hubbard
Hamiltonian we calculate the superconducting transition temperature $T_c$ as
a function of the hole content $n_h$. This method can be used to determine
the critical temperatures $T_c$ either for the $extended-s$ wave or $d$ wave
symmetry. We can also describe the pressure effects on $T_c$ by introducing
the assumption that it induces a change in the magnitude $V$ of the
attractive potential. This assumption yields an explanation for the intrinsic
term, and together with the well known change in $n_h$, we set the critical
temperature as $T_c=T_c(n_h(P),V(P))$. Then, we obtain a general expansion of
$T_c$ in terms of the pressure $P$ and the hole content $n_h$. We apply this
expansion to the $Tl_{0.5}Pb_{0.5}Sr_2Ca_{1-x}Y_xCu_2O_7$ system ($0.0\le x\le
0.35$) and our results provide a good fitting for the experimental data for the
variation of $T_c$ with the pressure $P$ and $n_h$.
\end{abstract}

\begin{keyword}
 BCS Model, Hubbard Hamiltonian, Phase Diagram, Pressure Effects
\end{keyword}
\end{frontmatter}
\newpage

\section{Introduction}

In spite of the advance attained in the
research of the high temperature superconductors (HTSC), a great number of
questions related to them remain to be answered. In order to obtain a better
understanding of these new materials many experiments were performed, but the
different experiments were interpreted by different theories. One of the most
fundamental evidence that these experiments showed was the interplay among
critical temperature $T_c$, hole content $n_h$, and pressure $P$, since one of
the effects of the application of external pressure is a change in the hole
content $n_h$ in the $CuO_2$ planes~\cite{30}, which results in a large change
in $T_c$. The large increase of $T_c$ by the application of pressure~\cite{9}
has been the motivation for a vast number of works on the study of the effects
of pressure on the critical temperature $T_c$ of the high temperature
superconductors~\cite{32,99,36,37}, not only to achieve higher $T_c$ but also
to obtain a hint for chemical substitution which originates new materials and
yields a better understanding of their physical properties. 

Thus, in the detailed study of the dependence of the critical temperature $T_c$
of the HTSC on the pressure $P$, it has been suggested that the
derivate $dT_c/dP$ may be strongly influenced by a pressure-induced alteration
in the density of hole-like charge carriers $n_h$ on the $CuO_2$
planes~\cite{29}. This proposal was also supported by experimental observations
of Hall constant's pressure dependence~\cite{30} that confirmed the existence
of a growth in $n_h$ in the $CuO_2$ planes when the pressure is
applied~\cite{huang}. Therefore, there is a pressure induced charge transfer
(PICT) from the block layers into the $CuO_2$ planes, that is, $n_h=n_h(P)$.
Since one expects $dT_c/dP=0$ at the optimum doping, and $T_c(n_{op})$
generally increases with $P$, it is well accepted that the pressure also
increases $T_c$ by a different mechanism of ``intrinsic"
origin~\cite{32,36,27}, and such intrinsic term is a characteristic of each
family of compounds. Therefore, we can separate these two effects by
considering~\cite{32,36,27,33}:      

\begin{equation} \frac{dT_c}{dP}=\frac{dT_c^i}{dP}+\frac{\partial T_c}
{\partial n_h}{\partial n_h\over \partial P}. \label{a} 
\end{equation} 
Here the first term, $dT_c^i/dP$, represents the intrinsic contribution, not
related to the charge transfer induced by pressure $\partial n_h/\partial P$.
Thus, the change of $T_c$ with the hole content is given by $\partial
T_c/\partial n_h$. 

In this paper we consider the intrinsic contribution as originated
from a variation of the magnitude $V$, of the attractive potential that forms
the Cooper pairs in the superconducting phase with the applied pressure $P$,
in such a way that $V=V(P)$. Similar proposition was previously made by
Angilella et al.~\cite{27} and de Mello et al~\cite{32,36}. As discussed in
Ref.~\cite{32,99}, as $V$ increases with the applied pressure, the
zero temperature superconductor gap also increases. This hypothesis has been
indirectly verified in the experimental study of polycristallines samples of
$Hg_{0.82}Re_{0.18}Ba_2Ca_2Cu_3O_{8+\delta}$~\cite{Gonzales}. 

\section{The Method} 

In order to study the dynamics of the hole-type carriers in the superconducting
phase as well as the normal phase with correlations and the basic attractive
interaction, we consider a two dimension extended Hubbard
Hamiltonian~\cite{32,36,27} 

\begin{eqnarray} 
H= -\sum_{\ll ij\gg \sigma}t_{ij}c_{i\sigma}^\dag
c_{j\sigma} +
U\sum_{i}n_{i\uparrow}n_{i\downarrow} 
+\sum_{<ij>\sigma
\sigma^{\prime}}V_{ij}c_{i\sigma}^\dag c_{j\sigma^{\prime}}^\dag
c_{j\sigma^{\prime}}c_{i\sigma}, \label{b}    
\end{eqnarray} 
where $t_{ij}$ is the hopping integral between the nearest-neighbour and
next-nearest-neighbour sites $i$ and $j$; $U$ is the on-site correlated
repulsion and $V_{ij}$ describes the attractive interaction between
nearest-neighbour sites $i$ and $j$.

In the superconducting phase the paired states form a condensate separated
from the single-particle states by a ``gap" (order parameter). Using a
BCS-type mean-field approximation~\cite{26} the self-consistent gap equation, at finite
temperatures, is given by

\begin{equation}
\Delta_{\bf k}=-\sum_{\bf k^{\prime}}V_{\bf
kk^{\prime}}\frac{\Delta_{\bf k^{\prime}}}{2E_{\bf
k^{\prime}}}\tanh\frac{E_{\bf k^{\prime}}}{2k_BT}, 
\label{c}
\end{equation}

with
\begin{equation}
E_{\bf k}=\sqrt{\varepsilon_{\bf k}^2+\Delta_{\bf k}^2}. \label{r}
\end{equation}
Here, $V_{\bf kk^{\prime}}$ is the interaction potential which contains the
repulsive and attractive potential, and may be written in a separable
form~\cite{27,Schneider}

\begin{equation}
V_{\bf kk^{\prime}}=U+2V\cos(k_xa)
\cos(k_x^{\prime}a)+2V\cos(k_ya) \cos(k_y^{\prime}a), \label{d}
\end{equation}
where it was considered a square lattice of lattice parameter $a$ and equal
coupling constants $V$ along both directions in the Cu-O planes, in the
context of mean field approximation. 

A tight-binding approximation may be employed to dispersion relation in
Eq.\ref{r}, which yields 

\begin{equation}
\varepsilon_{\bf k} = -2t(\cos(k_x a)+\cos(k_y
a))+4t_{xy}\cos(k_xa)\cos(k_ya)-\mu. \label{e}
\end{equation}
This equation has been previously considered~\cite{32,36,27,Schneider}, but
here it was considered identical hopping integrals along both directions in
the Cu-O planes for the nearest-neighbour ($t_x=t_y=t$) and a different one
for the next-nearest-neighbour ($t_{xy}$); $\mu$ is the chemical potential. 

Also, using  the same BCS-type mean-field
approximation applied to the gap (Eq.\ref{c}), we obtain the
hole-content equation~\cite{24}

\begin{equation} 
n_h(\mu,T)=\frac{1}{2}\sum_{\bf k}\left(
{1-\frac{\varepsilon_{\bf k}}{E_{\bf k}} \tanh\frac{E_{\bf k}}
{2k_BT}}\right), \label{f}
\end{equation}
where $0\leq n_h\leq 1$.

Following the steps of Ref~\cite{27}, one observes that the substitution of
the potential (Eq.\ref{d}) in the gap equation (Eq.\ref{c}) leads to
the appearance of a gap with two different symmetries  

\begin{equation}
\Delta_{\bf k}(\mu,T)=\Delta^{max}(\mu,T)[\cos (k_xa)\pm \cos (k_ya)],
\label{g} \end{equation}
where the plus sign is for $extended-s$ wave and the minus sign, for
$d$ wave symmetry. The anisotropy of the gap is one of the characteristic 
distinction between the usual low temperatures superconductors and the HTSC.
Many experiments and theoretical calculations suggest that the HTSC exhibit a
pairing symmetry different than the usual superconductors~\cite{18}, and in
some cases even a mixture of different symmetries must be
considered~\cite{mixture}. Thus, the determination of the symmetry of the gap
is the first step in the identification of the pairing mechanism of the charge
carriers and the subsequent development of a microscopic theory for the HTSC.

To obtain the phase diagram, the gap equations are
solved numerically, in the limit $T\longrightarrow T_c$, together and
self-consistently with the density of hole carriers $n_h$ (Eq.\ref{f}). In
these calculations it was considered that there were no symmetry mixture near
the critical point~\cite{27,mixture}.

The values of the hopping integrals $t$ and $t_{xy}$ used in the calculation
of the phase diagram of the $Tl_{0.5}Pb_{0.5}Sr_2Ca_{1-x}Y_xCu_2O_7$ series are
taken from A.R.P.$\\$E.S. measurements for the Bi2212~\cite{27}, since it is a
HTSC with similar structure, having two $CuO_2$ identical layers per unit
cell~\cite{16}, although this does not guarantee that the real $t$ and
$t_{xy}$ have this values and that is why we have tried a small variation
around the A.R.P.E.S results. The chemical potential $\mu$ is
calculated self-consistently. The values of $U$ and $V$ are considered
adjustable parameters, in order to reproduce the $T_c\times n_h$ curve close
to the experimental data. Our method to derive the phase diagrams is similar
to the approach based on a change in the zero temperature
gap~\cite{deMello2}. On Fig. 1 we show the phase diagram with the theoretical
curves from the numerical calculation for the $extended-s$ and $d$ wave
symmetries, together with the experimental data for the
$Tl_{0.5}Pb_{0.5}Sr_2Ca_{1-x}Y_xCu_2O_7$~\cite{19}.

\begin{figure}[!h] 
\epsfysize=\textwidth
\centerline{\epsfbox{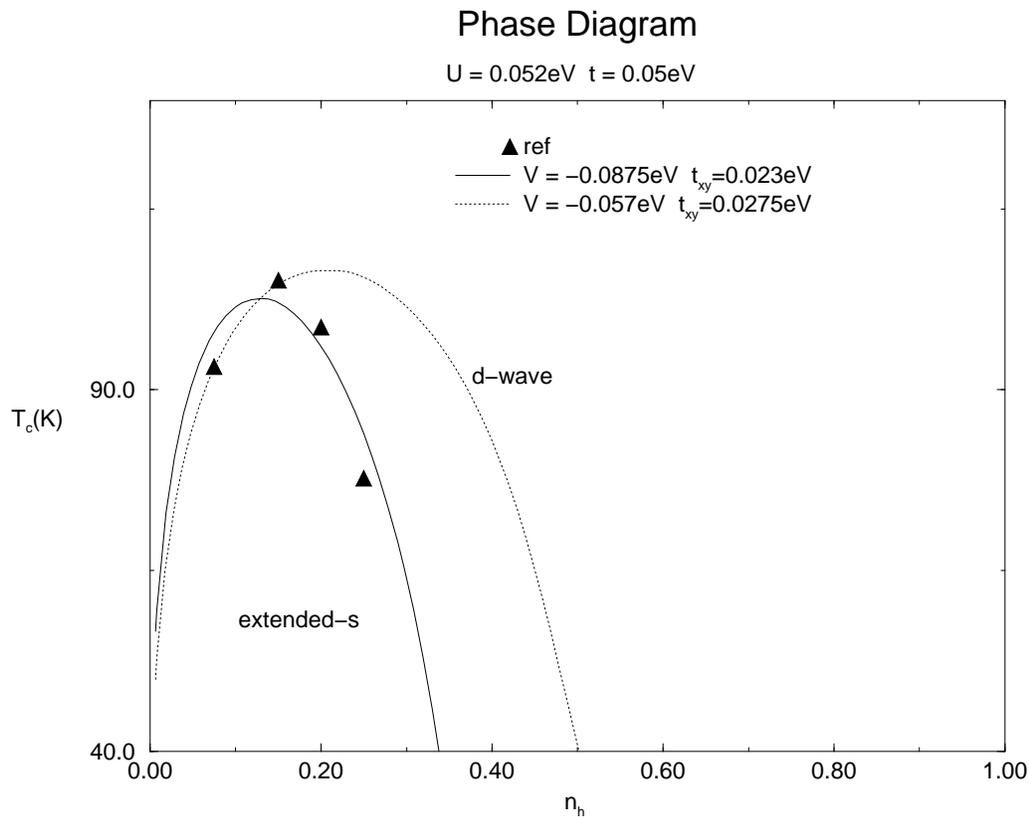}}
\caption{Phase diagram showing the numerical calculation for the
$extended-s$ and $d$ wave symmetry together with the experimental data of the
$Tl_{0.5}Pb_{0.5}Sr_2Ca_{1-x}Y_xCu_2O_7$ series taken from Ref.~\cite{19}.}
\end{figure}

Fig. 2 illustrates the behavior of the phase diagram, for the $extended-s$
symmetry, when $V$ is varied. Both symmetries exhibit a similar dependence on
the potential $V$.

\begin{figure}[!h] 
\epsfysize=\textwidth
\centerline{\epsfbox{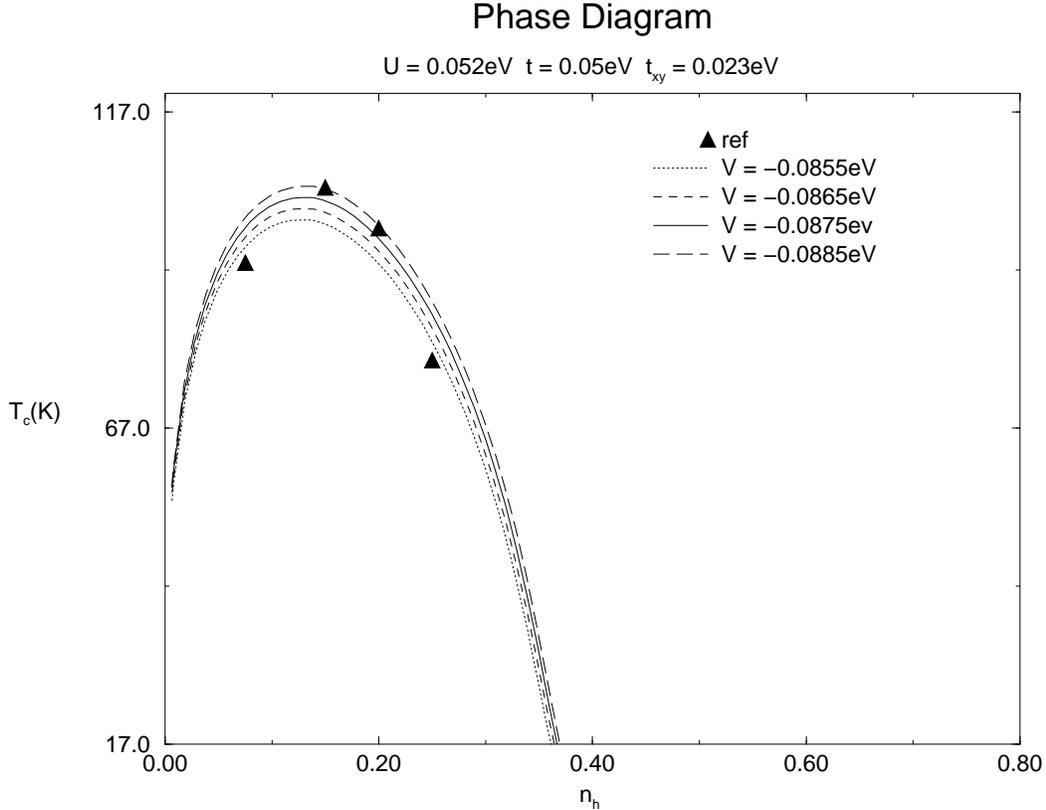}}
\caption{Phase diagram of the
$Tl_{0.5}Pb_{0.5}Sr_2Ca_{1-x}Y_xCu_2O_7$ series showing the effect of a change
in the magnitude $V$ of the attractive potential.}
\end{figure}

Notice that although there is a strong tendency toward a $d$-wave symmetry on
many experiments the HTSC, our results, with the parameters $t$ and $t_{xy}$
fixed, show that the $extended-s$ symmetry seems to reproduce well the phase
diagram of the $Tl_{0.5}Pb_{0.5}Sr_2Ca_{1-x}Y_xCu_2O_7$ series in a better
agreement with the experimental points than the $d$ wave one. This is
in accordance with the results of Schneider et al.~\cite{Schneider}. 

Thus, the study of the phase diagram yields the best values $t$=0.05eV,
$t_{xy}$=0.023eV, $U$=0.052eV and $V$=-0.0875eV for the $extended-s$
symmetry, and  $t$=0.05eV, $t_{xy}$=0.0275eV, $U$=0.052eV and $V$=-0.057eV for
the $d$ wave symmetry, which will be used in the next section.  

\section{ The Pressure Effects }

On Fig. 2 we show that $T_c$ is quite sensible with respect to a change
in the magnitude $V$ of the attractive potential. Therefore, as it was
previous mentioned, we suppose that the pressure induces a change in the
attractive potential, in such a way that $V=V(P)$. In the absence of
informations of how $V$ effective behaves with respect to the pressure
$P$, we assume a linear dependence~\cite{32,36,34}, which should be the case
since the pressure causes very small structural changes in the lattice
constants~\cite{27}. Therefore    

\begin{equation}
V(P)=V+\Delta V(P). \label{h}
\end{equation}
In the same way, for the density of carriers, we have
 
\begin{equation}
n_h(P)=n_h+\Delta n_h(P). \label{i}
\end{equation}
Here, $V$ and $n_h$ are the $P=0$ values of $V(P)$ and $n_h(P)$. The
pressure dependence is defined as 

\begin{equation}
\Delta V(P)=c_1P,\hspace{0.7cm} \Delta n_h(P)=c_2P \label{j}
\end{equation} 
where $c_1=({\partial V/\partial P})$ and $c_2={(\partial n_h/\partial
P)}$ are parameters usually determined from the experimental data. 

Our previous discussion leads to $T_c(n_h,P)=T_c(n_h(P),V(P)).$
Thus, we can estimate $T_c$ for a compound
with a certain value $n_h$, and under pressure $P$, using an expansion of
$T_c(n_h,P)$ in terms of $P$. Consequently, we have
   
\begin{eqnarray}
T_c(n_h,P)&=&T_c(n_h,0)+\left ( \frac{dT_c}{dP}\right
)_{P=0}P+\frac{1}{2!}\left ( \frac{d^2T_c}{dP^2}\right )_{P=0}P^2+ 
\nonumber \\ 
&&\frac{1}{3!}\left ( \frac{d^3T_c}{dP^3}\right )_{P=0}P^3+\cdot
\cdot \cdot, \label{k} 
\end{eqnarray}
where

\begin{equation}
{d^zT_c\over dP^z}=\left ( c_1{\partial \over
\partial V}+ c_2{\partial \over \partial n_h}\right
)^zT_c(n_h(P),V(P)). \label{l} 
\end{equation}
$T_c(n_h,0)$ is the critical temperature for $P$=0. The Eq.
(\ref{k}) can be written as  

\begin{equation}
T_c(n_h,P)=\sum_z\alpha_z{P^z\over z!}, \label{m} 
\end{equation}
with

\begin{equation}
\alpha_z=\left ( c_1{\partial \over \partial
V}+c_2{\partial \over \partial n_h}\right )^zT_c(n_h(P),V(P)). \label{n} 
\end{equation}
One can see by considering the case $z=1$ in the above equation, and comparing
with Eq.(\ref{a}), that

\begin{equation}
{dT_c^i\over dP}={\partial V\over \partial P}{\partial T_c\over \partial
V},\label{o}  
\end{equation}
which clearly shows that, in our model, the intrinsic contribution is related
to the variation of the magnitude $V$ of the attractive potential with the
pressure $P$. Now, the analytical expressions for the coefficients $\alpha_z$
will be derived below as a function of the parameters $c_1$ and $c_2$, for the
values of $n_h$ and $P$.

The first coefficient ($z=1$) for the above expression is given by 

\begin{equation}
\alpha_1=\left ( c_1{\partial T_c\over \partial V}+c_2{\partial T_c\over
\partial n_h}\right ). \label{p}  
\end{equation}
Restricting ourselves to small changes in $V$, we may approximate  

\begin{equation}
{\partial T_c\over \partial V}\approx \overline{\Delta T_c\over \Delta V},
\label{q} 
\end{equation}
where the horizontal bar denotes a ``mean" over the values of $T_c$ obtained
direct from the phase diagram as a function of $V$, as it is shown in Fig. 2.
This ``mean" is realized for each value of $n$. To estimate $\partial
T_c/\partial n_h$ we may either use an experimental value, or use directly the
curves of Fig. 2 or a phenomenological universal parabolic curve~\cite{19}: 

\begin{equation}
T_c=T_c^{max}[1-4\beta (n_{op}-n_h)^2],\label{q1}
\end{equation}
and, therefore, an analytical expression for $\partial T_c/\partial n_h$ is
obtained.  Thus, we derive the following expression for the coefficient
$\alpha_1$ 

\begin{equation}
\alpha_1=c_1\overline{\Delta T_c\over \Delta
V}+c_22(4\beta)T_c^{max}(n_{op}-n_h), \label{t}
\end{equation}
and $\alpha_2$, which is given by 

\begin{equation}
\alpha_2=\left ( c_1{\partial \over \partial V}+c_2{\partial \over \partial n_h}\right ) 
\left ( c_1{\partial T_c \over \partial V}+c_2{\partial T_c \over \partial
n_h}\right ), \label{u}
\end{equation}
and results in 

\begin{equation}
\alpha_2=2(4\beta )\overline{\Delta T_c^{max}\over \Delta V}(n_{op}-n_h)c_1c_2
-2(4\beta )T_c^{max}c_2^2,\label{v}
\end{equation}
where ($\overline{\Delta T_c^{max}\over \Delta V}$) is an empirical mean
derived numerically by inspection on Fig. 2 at the optimum doping compound
$n_{op}$. Following along the same procedure, the third coefficient becomes

\begin{equation} 
\alpha_3=-4(4\beta )\overline{\Delta T_c^{max}\over \Delta
V}c_1c_2^2.\label{x}
\end{equation}
As one can notice the coefficients $\alpha_z$, for $z\geq4$, are all nulls.

\section{Comparison with experimental data}

In order to study the effects of the pressure on the critical temperature of
any given family of compounds, we need first to calculate the constants $c_1$
and $c_2$, which yield the coefficients $\alpha_z$. Thus, we can see that the
constants $c_1$ and $c_2$ are determined by how $V$ and $n_h$ changes with the
pressure (Eq (\ref{j})). Hall effects can give directly $\partial n_h/\partial
P$ as in the case of Bi2212~\cite{huang} and therefore $c_2$~\cite{99,36,37}.
Since these measurements are not available for the
$Tl_{0.5}Pb_{0.5}Sr_2Ca_{1-x}Y_xCu_2O_7$, we can obtain $c_1$ and $c_2$ by
comparing with the curves of $T_c\times P$ for two values of $n_h$. In fact
Wijngaarden et al.~\cite{16} also estimate the value of $c_2$ from their
experimental data for the $Tl_{0.5}Pb_{0.5}Sr_2Ca_{1-x}Y_x$\\$Cu_2O_7$.

Since at low pressures only the linear term comes into
play, the higher order coefficients can be ignored. Thus, $\alpha_1$ becomes
the slope of a linear approximation from the initial points of the $T_c\times
P$ curve, for a given $n_h$ compound. Starting with $n_h$=$n_{op}$, in spite
of not being strictly necessary, we determine $c_1$ using the estimated
$\alpha_1$ in Eq.(\ref{t}), as long as at $n_{op}$ the charge transfer term
vanishes. To determine $c_2$ we estimate $\alpha_1$ from the $T_c\times P$
curve of an $n_h\not= n_{op}$ compound, and use again Eq.(\ref{t}). Once
these two constants are determined, the $\alpha_z$ coefficients for
any $n_h$ value can be calculated (Eq.(\ref{n})). 

For the $Tl_{0.5}Pb_{0.5}Sr_2Ca_{1-x}Y_xCu_2O_7$~\cite{16,19} series we used
the experimental values for the $n_h=n_{op}=0.15$ and $n_h$=0.20 compounds to
calculate $c_1$ and $c_2$. The estimated $\alpha_1$ were 1.5$K/GPa $ and
0.41$K/ GPa$ for $n_{op}$ and $n_h$=0.20, respectively. The phase diagram
parameters ($\overline{\Delta T_c\over \Delta V}$) taken from Fig. 1 and 2
and used in the calculations were 2900$K/eV$ for the $extended-s$ wave and
2300$K/ eV$ for the $d$ wave symmetry for the $n_{op}$ compound and 2700$K/eV$
for the $extended-s$ wave and 2600$K/ eV$ for the $d$ wave symmetry for the
$n_h$=0.20 compound. The parameters of the parabolic curve (Eq.\ref{q1}) used
were $\beta=3.5$ and $T_c^{max}$=105.1K. Therefore, the resulting constants
were $c_1^s=5.17\times 10^{-4}eV/GPa$ and $c_2^s=6.7\times10^{-3}GPa^{-1}$ for
the $extended-s$ wave and  $c_1^d=6.52\times10^{-4}eV/GPa$ and
$c_2^d=8.7\times10^{-3}GPa^{-1}$ for the $d$ wave symmetry. 

On Fig. 3 and 4 we present our results in comparison with the experimental
data. One observes that, for the optimum doping, and the compounds near the
optimum doping, we have a rather good result. For the optimum compound
($n_h$=0.15) in particular, we had the better result, for both symmetries, with
very good agreement with the experimental data~\cite{16,19}. For
$n_h$=0.075, 0.20 and 0.25 we can reproduce well the low temperature data and
obtain a qualitative agreement on the high pressure region.

\begin{figure}[!h] 
\epsfysize=\textwidth
\centerline{\epsfbox{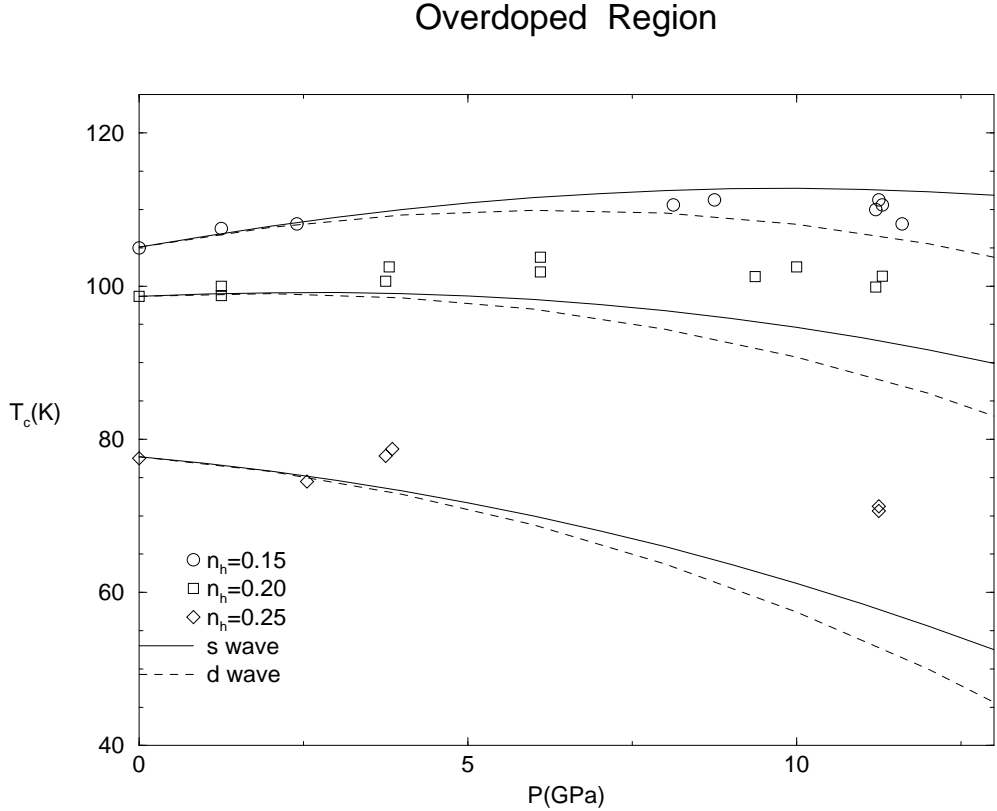}}
\caption{Numerical calculation for the $extended-s$ (filled
lines) and $d$ wave (dashed lines) symmetries, in the overdoped region,
together with the pressure experimental data of the
$Tl_{0.5}Pb_{0.5}Sr_2Ca_{1-x}Y_xCu_2O_7$ series taken from Ref.~\cite{19}.}
\end{figure}

\begin{figure}[!h] 
\epsfysize=\textwidth
\centerline{\epsfbox{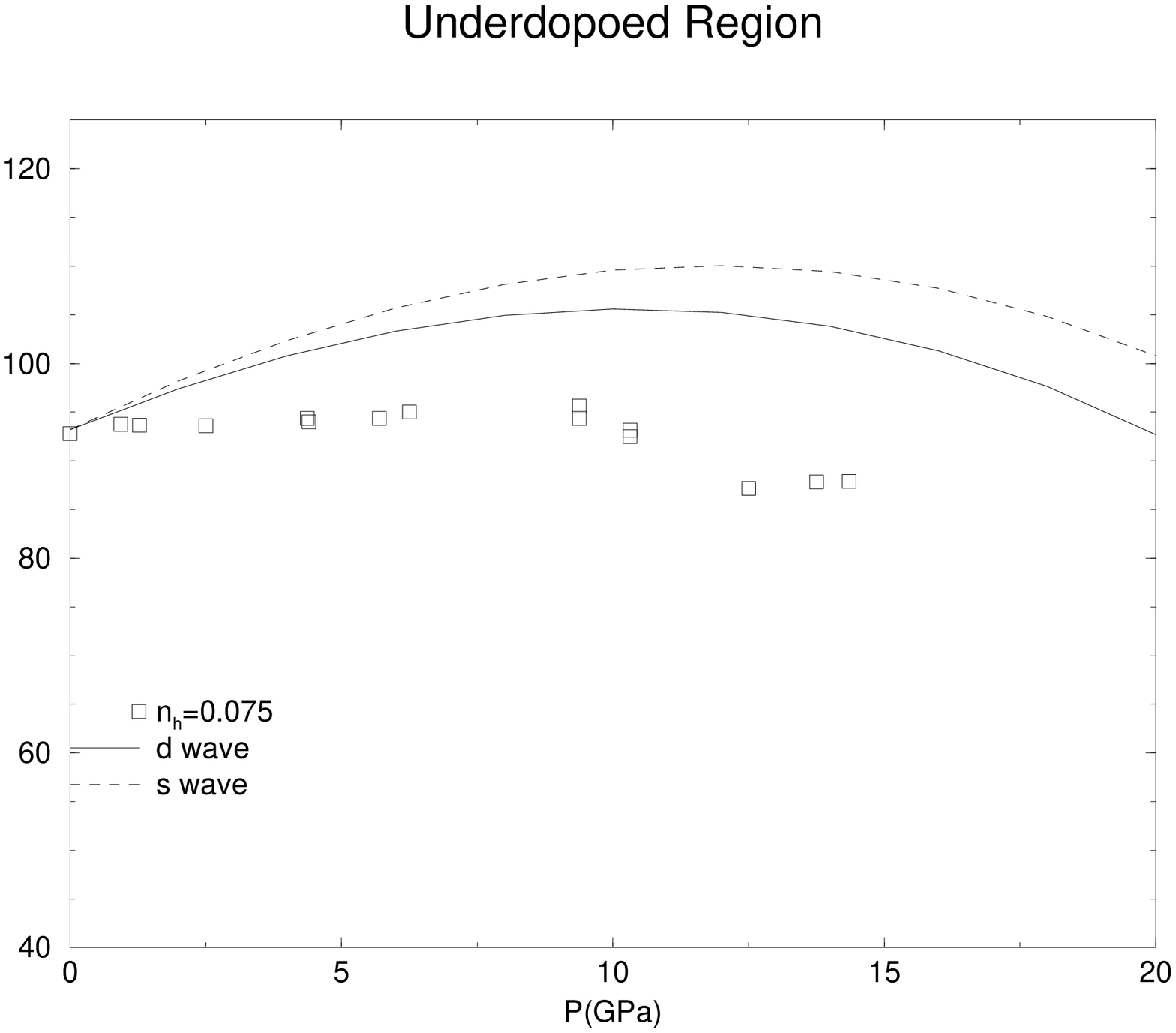}}
\caption{Numerical calculation for the $extended-s$ (filled
lines) and $d$ wave (dashed lines) symmetries, in the underdoped region,
together with the pressure experimental data of the
$Tl_{0.5}Pb_{0.5}Sr_2Ca_{1-x}Y_xCu_2O_7$ series taken from Ref.~\cite{19}}
\end{figure}

The value of the constant $c_2$ ($=6.7\times 10^{-3}GPa^{-1}$), which
means the rate of charge transfer by the pressure (${\partial n_h \over
\partial P}$), for the $extended-s$ symmetry is close to the experimental
results for the $YBCO$ compounds~\cite{33}, and is smaller than the
experimental Hall coefficient value for the Bi2212~\cite{huang}. The value
estimated experimentally by Wijngaarden et al.~\cite{16,19} for the
$Tl_{0.5}Pb_{0.5}Sr_2Ca_{1-x}Y_xCu_2$\\$O_7$ series was ${\partial n_h/
\partial P}_{exp}=3.9\times {10}^{-4}{GPa}^{-1}$. This discrepancy can be
attributed to the parameters $\overline{\Delta T_c\over \Delta V}$, which are
difficult to be estimated with the very few experimental points, what
introduces some uncertainties in the calculated phase diagrams. This
discrepancy is also the reason why our curves is in good agreement with the
experimental data up to $10GPa$, but starts to deviate above this value.
Another evidence that our main source of error comes from ${\partial n_h/
\partial P}$ is the fact that the optimum compound, with $n_h$=0.15, which is
the one with ${\partial n_h/\partial P}$=0, is in much better agreement with
the experimental results than the others, with different $n_h$.   

\section{Conclusions}

In conclusion, to relate the pressure $P$, the density of holes $n_h$ and the
attractive potential $V$ we used in the calculation a BCS-type mean-field
approach on the bidimensional extended Hubbard Hamiltonian with the potential
$V$ as an adjustable parameter. Our method has general application, and can
therefore be used in any compound under pressure. We showed that these
calculations can describe reasonably well the pressure effects on the critical
temperature $T_c$ for the series $Tl_{0.5}Pb_{0.5}Sr_2Ca_{1-x}Y_xCu_2O_7$, for
the four measured doping values.

Our results for the $extended-s$ wave symmetry give a better quantitative
agreement with the experimental data. But the $d$-wave calculations are also
in qualitative agreement with the experimental data, showing that the order
parameter symmetry is not decisive to study the pressure effects. 

Thus, our method provides a very interesting novel explanation to the
intrinsic term, which is related with the change in the potential $V$, in
agreement with a previous hypothesis that the pressure increases the zero
temperature superconducting gap~\cite{32,99} and with some experimental
results~\cite{Gonzales}.

\end{document}